\documentclass[twocolumn,amsmath,showpacs,amssymb,aps,nofootinbib,floatfix]{revtex4-1}
\usepackage{grffile}
\usepackage{epsfig}
\usepackage{latexsym}
\usepackage{color}
\usepackage{slashed}

\newcommand{\beq}{\begin{equation}}
\newcommand{\eeq}{\end{equation}}

\newcommand{\eqcomma}{,\phantom{AA}}
\newcommand{\order}[1]{ \mathcal{O} \left( #1 \right) }
\newcommand{\ave}[1]{\left\langle #1 \right\rangle}

\begin{document}

\title{A Lagrangian formulation of relativistic Israel-Stewart hydrodynamics}

\author{David Montenegro, Giorgio Torrieri}
\affiliation{IFGW, Universidade Estadual de Campinas, Campinas, S$\tilde{a}$o Paulo, Brazil}

\date{\today}

\begin{abstract}
We rederive relativistic hydrodynamics as a Lagrangian effective theory using the doubled coordinates technique, allowing us to include dissipative terms.   We include Navier-Stokes shear and bulk terms, as well as Israel-Stewart relaxation time terms, within this formalism.  We show how the inclusion of shear dissipation forces the inclusion of the Israel-Stewart term into the theory, thereby providing an additional justification for the form of this term.
\end{abstract}

\pacs{25.75.-q,25.75.Dw,25.75.Nq}

\maketitle

\section{Introduction \label{secintro}}
Hydrodynamics, and its relativistic incarnation, is a topic of very active theoretical and phenomenological development \cite{kodama}.  Phenomenologically, it seems to provide a good description of physics in heavy ion collisions, making numerical hydrodynamic solvers an indispensable tool in this field.
These phenomenological applications are however hampered by the fact that a rigorous link between hydrodynamics and ``microscopic theory'' (in this case, Quantum Chromodynamics close to deconfinement temperature) is still missing. In fact, a general link between hydrodynamics and {\em any} microscopic theory, including models where hydrodynamic behavior appears most naturally as an infrared limit \cite{4pi}, is a surprisingly non-trivial problem.

There are several aspects to this: Extending Navier-Stokes hydrodynamics to the relativistic limit will give problems with causality \cite{israelst,romdev,romconf}, due to the presence of arbitrarily-high speed dissipative modes.
These can be removed by promoting all elements of the energy-momentum tensor to degrees of freedom, and giving them a Maxwell-Cattaneo type equation of motion \cite{israelst,romdev} characterized by a ``relaxation time'' transport coefficient.

Phenomenologically this works, and also avoids unphysical instabilities in the linearized expansion \cite{hiscock,shipu,marrochio} (such instabilities would affect the hydrostatic limit, making local entropy non-decrease problematic), but, if one regards the microscopic origin of hydrodynamics as a ``gradient expansion'' in terms of bulk quantities, it is not immediately clear why additional degrees of freedom need to arise at second order, nor what the relationship between relaxation time $\tau_\pi$ and the usual microscopic ``dimensionless small parameter'' (the Knudsen number) of the theory is.  One expects $\tau_\pi$ to scale as the sound wave attenuation length, but this is not universally true and in any case the extra degrees of freedom are determined by initial conditions, and there is no limitation to their size \cite{mauricio,mepratt}.   
The existence of this extra small parameter can also be understood from the form of the microscopic spectral function \cite{denicol}, but this turns out to be highly dependent on the exact microscopic theory we are dealing with.    Some works \cite{reynolds} consider it a second independent small parameter (``inverse Reynolds number'') to the gradient ``Knudsen number'' expansion.

In general, 
extending these approaches into a consistent general systematic small parameter expansion, with possibly additional higher-order tensor degrees of freedom, hundreds of higher order gradient terms,  and the requirement of overall hydrostatic stability entropy non-decrease seems only feasible for a select number of highly symmetric theories and boundary conditions \cite{series1,series2,series3}{.}

Microscopic fluctuation terms, most likely highly relevant in the low viscosity limit \cite{mooresound,kovfluct,betz,mevisc} are even more mysterious.
So far{,} they have been studied systematically only at linear order \cite{linfluct1,linfluct2}, but evidence exists \cite{hydro1,burch} they cannot at all be captured in a perturbative expansion.

A recent attempt to investigate this problem is to rewrite hydrodynamics as a field theory \cite{hydro0,hydro2,ritz1,grozdanov,galley,glorioso,floerchinger}, with the fields representing the Lagrangian coordinates of the fluid's volume elements. This picture allows the use  effective field theory techniques \cite{kaplan} to investigate links between microscopic and macroscopic theories without explicitly writing down the microscopic dynamics.  In this approach, hydrodynamics can be thought of as an example of an effective field theory with a cutoff, since we do track macroscopic degrees of freedom, but only conserved currents (energy-momentum, conserved charges) averaged over each fluid cell. Thus, long wave length and microscopic variables can be reformulated in the sense of effective field theory (EFT) as, respectively, infrared and ultraviolet degrees of freedom \cite{kaplan,betz}, the latter relevant at a microscopic scale $l_{UV}$ (either a small distance or a high wavenumber/momentum). In the case of hydrodynamics the Knudsen number provides a natural expansion parameter, combining the mean free path $l_{mfp}$ and the gradients into a dimensionless parameter.
Effective Lagrangian terms are calculated or correcting with systematic expansion of higher order derivatives  $(l_{UV}. \vec{\nabla})^n$ and the fundamental symmetries of the system.

The big apparent limitation of such an approach is that leading corrections in hydrodynamics are dissipative, and standard Lagrangian theory cannot deal with them.   However, methods have been developed to overcome this problem \cite{caltech,morse,galley1,porto}.   Provided these can be developed consistently for all orders in the EFT, the problems highlighted in the preceding paragraph are all solvable systematically.  EFT techniques will automatically separate physical from non-physical terms, and fluctuations can be included without any approximations by promoting the least action trajectory into a functional integral \cite{hydro1,mevisc,burch}.
\begin{figure}[t]
  \includegraphics[height=0.2\textheight]{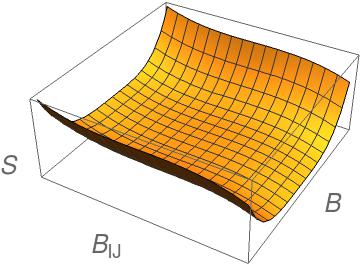}
  \includegraphics[height=0.2\textheight]{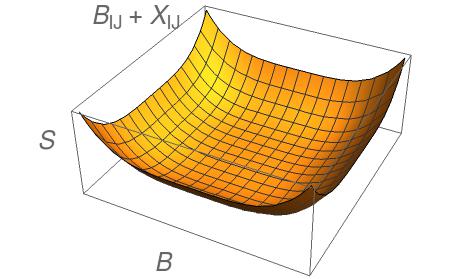}
  \caption{\label{action}The qualitative form of the action $S$ in terms of the $b$ {$B$} and $B_{IJ}$ degrees of freedom for Navier-Stokes (top panel) and Israel Stewart hydrodynamics (bottom panel) close to the hydrostatic ($\phi^I \simeq x^I$) limit.  The first exhibits a saddle point, the other is positive definite.
}
\end{figure}
This works in this direction using the ``doubled coordinate approach'' outlined in Appendix \ref{appctp} \cite{galley,grozdanov} ({in this work, a variable with a subscript $\pm$ is doubled, without the subscript it is standard non-dissipative}).
As this work will show, the appearance of extra degrees of freedom at second order, and the relationship between the relaxation time and the microscopic scale, also look natural within the Lagrangian formalism once the existence of semi-classical Lagrangian and ``vacuum stability'' (the existence of an action extremum) are taken into account. We show that, as hypothesized in \cite{grozdanov}, the extra degrees of freedom appear already in the Navier-Stokes limit, but in a way that cannot give a stable hydrostatic vacuum. The extra degrees of freedom in Israel-Stewart \cite{israelst,romdev} are then required to stabilize the theory.

{More specifically, we shall show that the form of the Lagrangian for viscous hydrodynamics has to be
{
\begin{equation}\label{Hyexpansion}
\mathcal{L} = \mathcal{L}_{ideal}(B_{\pm}) + \mathcal{L}_{bulk}(B_{\pm},u^\mu_{\pm}) +{ \mathcal{L}_{shear}(B_{\pm}, u^\mu_{\pm},B_{IJ\pm})}
\end{equation}}
The term $\mathcal{L}_{ideal}(B)$  is the one studied in \cite{hydro0,hydro1,hydro2}, corresponds to a conserved local quantity (the microscopic entropy. An infinite number of non-local conserved vorticies are also present \cite{hydro0}) and generates Euler's equations. $\mathcal{L}_{bulk}$ does not require extra degrees of freedom, but needs to generate dissipative terms, something done here using the doubling of DoFs described in appendix  \ref{appctp}.
Shear viscosity requires, as well as doubling,  new terms ( $B_{IJ}$) which not correspond to a conserved quantity and break volume-preserving diffeomorphisms, and yet are necessary to define the shear viscosity. Mathematically, they represent the dissipation of macroscopic energy by microscopic degrees of freedom, above the cutoff.}

As we will show, extra degrees of freedom appear in $\mathcal{L}_{shear}$ ``implicitly'', without touching the equations of motion.  As long known \cite{caltech}, actions based on such lagrangians are unstable, without global minimum (Fig. \ref{action} top panel). {\em this can be understood as the fundamental reason for the instability of the Navier-Stokes equations}.   Furthermore, as we shall show the only way to stabilize this system (have a Lagrangian of the form Fig. \ref{action} bottom panel) is to {modify $\mathcal{L}_{shear,bulk}$ into} a term of the form $\mathcal{L}_{IS}$, with additional degrees of freedom $X_{IJ}$ (corresponding to shear stress terms  {$\Pi_{\mu \nu}$ in the comoving frame, now} promoted to independent degrees of freedom), which gives rise to Israel-Stewart type dynamics, {The Lagrangian of this new dynamics is therefore of the form
{
\begin{equation}\label{Hyexpansion2}
\mathcal{L} = \mathcal{L}_{ideal}(B_{\pm}) +  { \mathcal{L}_{IS}(B_{\pm},u^\mu_{\pm},X_{IJ\pm},B_{IJ\pm}).} 
\end{equation}}
\section{ \label{secideal}A review of ideal hydrodynamics}
In this section, we give a review of the understanding of Lagrangian hydrodynamics as a field theory, developed in \cite{hydro0,hydro1,mevisc,burch}{.}

Let us consider an uncharged fluid element. {In} in the Lagrangian formulation of hydrodynamics it can be characterized by three scalar fields  $\phi^I(\vec{x},t)$ as d.of., where $\phi^I$ is simply the Lagrangian coordinate of the comoving volume, and $\vec{x}$ the Eulerian (lab) spacetime coordinate.  At thermostatic equilibrium these coordinates coincide so  $\ave{\phi^I} = x^I$, with  $I=\left\{ 1,2,3 \right \} $, it identify physical coordinate in flat $3+1$ dimensional with metric $g_{\mu\nu}=diag(-1,1,1,1)$\footnote{{Greek letters are used here to specify spacetime coordinates, while latin letters specify comoving coordinates.  The Einstein summation convention, with respectively 4D Minkowski and 3D Euclidean metrics, is used unless specified} } \cite{hydro1}.

In this formalism the dynamics of fluids, as opposed to other continuous media, can be built up by imposing symmetries on the Lagrangian.  
 Translation symmetry ($\phi^{I } \rightarrow \phi^{I }+a^I$ with $ a^I = const$ ) forces the Lagrangian to depend on derivatives of $ \phi^{I } $. Rotation symmetry (isotropy, $ \phi^{I } \rightarrow R^I_{J}\ \phi^{J }), \ \ R^I_J \in SO(3)$ is straightforwardly  implemented by requiring the Lagrangian to be a function of $ B_{IJ}$. {Volume preserving deformation symmetry ($\phi^{I} \rightarrow \xi^{  I}(\phi^{J })$, where $\xi^I$ is a different set of coordinates with the same volume element ( i.e. det$(\partial \xi^{I}/\partial\phi^{J})=1$) implies the Lagrangian depends solely on B}.   (in the hydrostatic limit each such symmetry gives the "Goldstone bosons" of the sound waves) \cite{hydro2}.   

In summary, an ideal fluid is described as
{
\begin{equation}\label{idealhy}
\mathcal{L}= T^4_0 F(B), \ \ \  B  \equiv  det \left | B^{IJ} \right | =  det \left | \partial_\mu \phi^{I} \partial^\mu \phi^{J}\right | {,}
\end{equation}}
{Where $T_0$ is a microscopic scale whose necessity is clear since $\partial \phi^{I}$ is dimensionless.   The role of this microscopic scale , (which also absorbs a microscopic degeneracy, such as $N_c^2$ in gauge theories) is extensively discussed in \cite{mevisc,burch} and will be discussed later in this work.}

It is straightforward derive the stress energy tensor from the (\ref{idealhy}) via the usual Noether current 
{
\begin{equation}
\label{noetherT}
 T_{\mu\nu} = \sum_{I} \frac{\partial \mathcal{L}}{\partial (\partial^\mu \phi^I)} (\partial_\nu \phi^I) - g_{\mu \nu} \mathcal{L} {,}
\end{equation}}
we get
{
\begin{equation}\label{idealtensor}
T_{\mu\nu}= T_0^4 \left( 2B\frac{dF}{dB} B^{-1}_{IJ}  A^{IJ}_{\mu \nu} - F g_{\mu\nu} \right) .
\end{equation}}
Where $ B^{-1}_{IJ}$ is inverse of matrix defined as 
\begin{equation}
  \label{BAdef}
B^{IJ} \equiv \partial_\mu \phi^I \partial^\mu \phi^J, \ \ \ \  A^{IJ}_{\mu \nu} \equiv \partial_\mu \phi^I \partial_\nu \phi^J {,}
\end{equation}

the average value is $\Delta^{\mu\nu} \equiv \langle  B^{-1}_{IJ} A^{IJ} \rangle  =  u^\mu u^\nu +  g^{\mu \nu}$ \cite{burch}. The tensor can be written in the usual hydrodynamic form
{
\begin{equation}\label{tensor}
T^{\mu\nu} = e \ u^\mu u^\nu + p \Delta^{\mu\nu}{,}
\end{equation}}
provided the fluid energy density and pressure are, respectively,
{
\begin{equation}
\label{rhopdef}
eT_0^{-4} = - F(B), \ \ \ \ \ \ \ pT_0^{-4}= F(B) - 2B \frac{dF}{dB}{,}. 
\end{equation}}
and the velocity is defined, via {$u^\mu \partial_\mu \phi^I =0$} $  \forall I,u_\mu u^\mu=-1$ as 
\begin{equation}
\label{flowdef}
u^\mu = \frac{1}{6 \sqrt B} \epsilon^{\mu \alpha \beta \gamma} \epsilon_{I J K} \partial_\alpha \phi^I  \partial_\beta \phi^J  \partial_\gamma \phi^K {.} 
\end{equation}
Using the Gibbs-Duhem relation, relating entropy $s$ to pressure $P$, energy density $\rho$ (Eq. \ref{rhopdef}) and temperature $T$
{
\begin{equation}
\label{gibbsduhem}
s =\left. \frac{dP}{dT}\right|_{V} = \frac{P + e }{T}{,}
\end{equation}}
 we obtain the thermodynamic parameters
{
\begin{equation}
s=T^3_o \sqrt B , \ \ \ \ \  T= {-} T_o  \frac{\sqrt{B}  dF/dB}{g}{,}
\end{equation}
(where to define temperature one needs to separate the microscopic degeneracy $g$ from $T_0$ \cite{burch} because of the heat capacity's explicit dependence on $g$)}

This entropy is the only locally conserved quantity (there are infinite numbers of non-local conserved vorticity charges \cite{landau} corresponding to the Noether charges of diffeomorphisms \cite{hydro0}), giving rise to the conserved current used in \cite{grozdanov}
\begin{equation}\label{entropy}
K^\mu = \sqrt B u^\mu {,}
\end{equation}
One can understand the above by {\em defining} {$u^\mu$} to be parallel to entropy flow (the so-called Landau frame \cite{landau}), since the Euler-Lagrange equations applied to the Lagrangian in Eq \ref{idealhy} will just yield $\partial_\mu K^\mu=0$.  The symmetry of this Lagrangian against deformation and rotation will then yield the fact that $u^\mu$ is always perpendicular to the gradients of $\phi^I$.

{This way one can always construct $K^\mu$ and $u^\mu$ out of $\phi^I$ via a projector
\textcolor{black}{
\begin{equation}
\label{projector2}
P^{\mu \nu}_K = \frac{1}{3!} \epsilon^{\mu \alpha \beta \nu} \epsilon_{IJK} \partial_\alpha \phi^I \partial_\beta \phi^J,
\end{equation}}
This definition is in accordance with Eq. \ref{entropy} since clearly
\begin{equation}\label{defk}
 K^\mu \equiv P^{\mu \nu}_K \partial_\nu \phi^K.
\end{equation}}

\section{Navier-Stokes \label{secns}}
{The natural first order term within the Lagrangian is first order, i.e. contains exactly one gradient.   Such terms, however, are, 
as shown in \cite{hydro2}, non-dynamical since they can always be reabsorbed into field redefinitions.    Physically, this should not surprise us since we know that first order terms in the Navier-Stokes equations are dissipative, and hence cannot be represented by ``normal'' Lagrangian terms.
Such terms are{,} however{,} amenable to be included via the doubled degrees of freedom formalism \cite{caltech,morse,galley1,galley} described in appendix \ref{appctp}.
In this approach, the two degrees of freedom (represented here by $\phi_{\pm}$ according to the formalism introduced within \cite{grozdanov}) can be taken as representing the ``system'' and ``unobserved environment'',with the two distinguished by advanced and retarded boundary conditions \cite{galley}. }

Coarse-graining usually implies that the dynamics of the theory is a function of currents of conserved quantities, since these are the slowest to equilibrate.  These are related to symmetries by Noether's theorem, as well as their derivatives. Since in the last section, we found that the only conserved local charge is the entropy. We construct two independent vectors by doubling the degrees of freedom (DoFs) of Eq. \ref{entropy}, {we get two currents degenerated and independent}
{
\begin{equation}\label{kctp}
 K^{i\mu} = \frac{1}{3!} \epsilon^{\mu\alpha_1 \alpha_2 \alpha_3}  \epsilon_{IJK}\partial_{ \alpha_1} \phi^{\sigma_1 I}\partial_{ \alpha_2}\phi^{\sigma_2 J}\partial_{ \alpha_3}\phi^{\sigma_3 K} {,}
\end{equation}}
{where} $(\sigma _1 \sigma _2 \sigma _3) = \left \{(---),(+++)\right \}$, for i=$\left \{0,3 \right \}$. The current vectors are still conserved by $\partial_\mu K^{i \mu} = 0$, with further combinations are in fact forbidden by conservation laws \cite{grozdanov}.
$K^\mu$ can be understood as a generalization of the entropy current with doubled fields (i=$\left \{0,3 \right \}$) \textcolor{black}{with the two new projectors constructed using the proprieties of Eq \ref{projector2} and \ref{defk}. } 

\textcolor{black}{
\begin{equation}\label{proj1}
P^{i \mu \nu}_K \partial^\lambda \phi^K = \frac{1}{3}(K^{i\mu} \Delta^{\nu \lambda} - K^{i\nu} \Delta^{\mu \lambda}){,}
\end{equation}
\begin{equation}\label{proj2}
P^{i \mu \nu}_K \partial^\lambda \partial_\nu \phi^K = \frac{1}{3} \partial^\lambda K^{i\mu}.
\end{equation}}

We use the CTP formalism to obtain the Lagrangian and extend the hydrodynamics formulation to first order terms.  Following the correction to Noether's theorem described in \cite{galley1} and the procedure of \cite{grozdanov}( Eq. \ref{ctp} in Appendix \ref{appctp}.).
We see that the stress-energy tensor in CTP can be obtained by varying just one field by the Noether current
\begin{equation}\label{phi}
\phi^+(x) \rightarrow \phi^+(x + a(x)){,} \ \ \ \  ,  \ \ \ \ \  \phi^-(x) \rightarrow   \phi^-(x).
\end{equation}
{Note} that in physical limit (p.l.), the $\phi^{+K}=\phi^{-K}$ involve $K^{i \mu}$=$K^\mu$ which also defines $K^{3\mu}\equiv P^{3 \mu\alpha}_K \partial_\alpha \phi^{+K}$ and $ P^{0\mu \alpha}_{K} \equiv 0 $.  As we deal with only variations in $\delta^{+}_\phi K^{i\mu} = iP^{i \mu \alpha}_K \partial_\alpha \delta \phi^{+K}$, the other one $\delta^{-}_{\phi} K^{i\mu} = 0$, after a shift the metric $x \rightarrow x + a(x)$, we get $\delta^{+}_x \phi^{+K} = a^\mu \partial_\mu \phi^{+K}$. By inspection its gives
\begin{equation}
\delta^+_x K^{i\mu}  =iP^{i\mu\alpha}_K (\partial_\alpha a_\lambda\partial^\lambda\phi^{+K}+a_\lambda\partial^\lambda\partial_\alpha\phi^{+K}).
\end{equation}

The displacement $a(x)$, one can identify a equation of motion for
$T_{\mu \nu}$ with a non-vanishing divergence which however depends on
second-order gradients \cite{grozdanov}.  

{Without dissipative terms, $+$ and $-$ Lagrangians are symmetric, hence the results of the previous section merely get doubled \cite{grozdanov}.  The Lagrangian of Eq. \ref{idealhy}, written in terms of {$K^{i\mu}$}, is then promoted to }
\begin{equation}
\mathcal{L}^{(0)}_{CTP} =  T^4_oF(K^3_\gamma K ^{3\gamma}) -  T^4_oF(K^0_\gamma K ^{0\gamma}){,}.
\end{equation}
{where clearly the first term can be regarded, in the notation of the previous section, as the equation of state defined by $B_+$ and the second term as the equation of state defined by $B_-$.  In this form, the lagrangian is just two doubled lagrangians for two fluids, not talking to each other.}

{Additional} terms, are however possible.  In \cite{grozdanov} the bulk viscosity term was constructed to be
\begin{equation}\label{bulkgroz}
\mathcal{L}^{(1)}_{CTP} = T^4_o \sum_{i,j,k}  z_{ijk}(K^{l\gamma}K^{m}_\gamma) K^{i \mu} K^{j \nu} \partial_\mu K^k_\nu.
\end{equation}
which contains the same symmetries of ideal hydrodynamics.
{Small Latin index (symbolizing where we are in the doubled lagrangian) are always summed over $\left \{0,3\right \}$. The $l,m$ is also a summation term, but, unlike non-dissipative terms, it mixes $0$ and $3$, because $l \ne m$ and $i,j,k$ is summed over (for additional details on this notation see \cite{grozdanov} section IV)}{.}

The eight coefficients $z_{ijk}$ reduce to four $\bar{z}_{ijk} \equiv \left.z_{ijk}\right|_{|\phi^{+K} = \phi^{-K}}$ due to CTP symmetry as well as positivity ( ensuring the system is dissipative rather than anti-dissipative, where gradients {\em grow}), see Eq. \ref{ctpdef}.
 These terms specify the dependence of the shear and bulk viscosity on entropy $B^{1/2}$ by contractions of $(K^{l}_{\gamma}K^{m\gamma}={B} )$ in the argument of coefficient $z_{ijk}$.
  
  As correctly noted in \cite{grozdanov}, the construction of the shear viscosity term is complicated by the fact that it breaks the volume preserving diffeomorphism symmetry, and hence cannot be just a function of $K$.  Noting that $B_{IJ}$ is the leading term allowed by all but the volume preserving symmetries, and any thermodynamic quantity can be represented as a function of $B$ only, and using
\begin{equation}
\label{projector}
u_\gamma \partial_\mu (\Delta^{\gamma \nu}) = \partial_\mu u_\nu =- u^\gamma B^{-1}_{IJ} \partial_\mu \partial_\gamma \phi^I \partial_\nu \phi^J {,}
\end{equation}
(easily proven by taking the hydrodynamic derivative of eq. (28) of \cite{hydro0} )   we arrive at
\begin{equation}\label{ns}
\mathcal{L}^{(1)}_{CTP} = T^4_o \sum_{i,j,k}  z'_{ijk}(K^{l\gamma}K^{m}_\gamma) B B^{-1}_{IJ}\partial^{\mu} \phi^{iI} \partial^{\nu} \phi^{jJ} \partial_\mu K^k_\nu {,}
\end{equation}
will, when converted to $\Delta T_{\mu \nu}$ via Eq. \ref{noetherT}, produce the usual first order Navier-Stokes' equation shear viscosity term $\eta$ and bulk viscosity $\zeta$ that represent most general possible correction in first order with the symmetry.  
\begin{equation}
\label{eqns}
\sigma^{\mu\nu} \equiv \eta \Delta^{\mu\alpha} \Delta^{\nu \beta}\left( \partial_\alpha u_\beta + \partial_\beta u_\alpha  - \frac{2}{3}\eta_{\alpha \beta} \partial_\lambda u^\lambda \right) +\zeta \Delta_{\mu \nu} \partial_\alpha u^\alpha {.}
\end{equation}
  
A similar lagrangian to Eq. \ref{ns} was recently derived in \cite{floerchinger} (section 7) using laboratory coordinates.

Summarizing the shear viscosity appears when we introduce ``transverse stress'' d.o.f.s to break rotational diffeomorphism symmetry of the Lagrangian, $B^{-1}_{IJ}\partial^\mu \phi^I \partial^\nu \phi^J$, where it doesn't volume-preserve diffeomorphisms by a transformation det$(\partial \xi^{+I}/\partial\phi^{+I})=1$. The bulk viscosity, on the other hand, can be formulated by a projection parallel to $K^\mu$.  As a result, the Noether current for diffeomorphism invariance, vorticity \cite{hydro0} is explicitly broken by Eq. \ref{ns}, something we know well since the presence of shear viscosity dissipates vorticity.   Bulk viscosity also violates vorticity conservation, but it does so via the ``source term'' of Noether's theorem for dissipative theories \cite{galley}{.}

The relation between the viscosity as usually defined and the matrix coefficients {as well as the instantaneus entropy $B$ is \footnote{{During a time $dt$, entropy should increase from $B$ to $B+dB$, where $dB/dt$ is a function of the gradient.  Viscosity in this context should be a function of $B$ only and this can generate ambiguity in the definition of viscosity,  See the discussion in \cite{grozdanov} around eq. 85,98,99.   This issue is also discussed in the conclusion section of this work.}}} ({note the mixing between shear and bulk viscosity terms and $z$})
\begin{eqnarray}\label{bulkeq}
\zeta & = & - B^{3/2}(\bar{z'}_{003} + \bar{z'}_{303} + 2\bar{z'}_{333} + 2\bar{z'}_{033} + \bar{z}_{333} + \bar{z}_{300} \nonumber \\
& & - \bar{z}_{303} + 3\bar{z}_{330}) + B^{5/2} (\bar{z'}_{003,03} +\bar{z'}_{303,03}+\bar{z'}_{333,03} \nonumber \\
& & +\bar{z'}_{033,03} -2 \bar{z}_{333,03} -2\bar{z}_{303,03} +2 \bar{z}_{330,03} +2 \bar{z}_{300,03}) \nonumber \\
& & - 4B^{5/2}(\bar{z}_{333,00} + \bar{z_{303,00}}) + 4B^{5/2}(\bar{z'}_{003,33} + \bar{z'}_{303,33}  \nonumber \\
& & + \bar{z'}_{333,33} + \bar{z'}_{033,33} + \bar{z}_{330,33} + \bar{z}_{300,33}),
\end{eqnarray}
\begin{eqnarray}
\eta = B^{3/2}(\bar{z}_{003} + \bar{z}_{033} + \bar{z}_{303} + \bar{z}_{333}).
\end{eqnarray}
Where $z_{ijk,lm} \equiv \partial z / \partial (K^l_\gamma K^{m\gamma})$. Eq. \ref{ns} is very strange since a term $B_{IJ}$, independent from $K^\mu$ and $B$, is present at linear order.   This term represents the dynamics of energy-momentum components perpendicular to flow, and cannot be put in terms of DoFs invariant under volume-preserving diffeomorphisms. 
Physically, the fact that the shear viscosity term also breaks volume-preserving invariance, something physically understandable since this kind of dissipation necessarily requires degrees of freedom moving from from IR to UV ``macroscopic'' to ``microscopic'' across a cutoff $ \sim \frac{1}{T_o} $, and the microscopic scale is expected to be immune to deformation.
The doubled degrees of freedom make it possible that these first order terms are physical rather than redundancy terms (as first order terms usually are \cite{caltech,hydro2}).  

This means, however, that the doubled Lagrangian does not have a minimum but at most a saddle point (Fig. \ref{action} top panel).  This means that while equations of motion can be constructed independent of $B_{IJ}$, these will be unstable against perturbations in the $B_{IJ}$ direction.   This, in fact, explains, in terms of the Lagrangian, the linear-order results of \cite{hiscock,marrochio,shipu} about the instability of Navier-Stokes hydrodynamics and makes it apparent non-linearities cannot cure the instabilities encountered in these works, at least at the classical level.

In reality, this condition exists implicitly for bulk viscosity as well, for while $K_\mu K^\mu$ is positive-definite, $K_\mu K_\nu \partial^\nu K^\mu$ is unbounded above or below (it actually generically follows shear viscosity in the sense that both $B_{IJ}$ and $K_\mu K_\nu \partial^\nu K^\mu$ can become large and arbitrarily positive/negative when gradients do not follow the direction of entropy flow ).  Hence, Eq. \ref{bulkgroz} is {\em also} unbounded below, as expected from the fact that bulk and shear viscosity instabilities arise in a very similar way \cite{shipu}  {.}
The way to stabilize the Lagrangian is clearly to add higher order well behaved (even-power) terms, something that requires a few subtleties, as explained in the next section.

\section{Israel-Stewart \label{secis}}

The NS shows unphysical behavior for short wavelength, with a causality problem is most clearly seen by considering linearized perturbations.  The equations of motion for a Lagrangian perturbed from the hydrostatic limit
\[\  \phi^I = x^I + \delta \pi^I(x^\mu) \eqcomma B \rightarrow B_0+\delta B(x^\mu)  \]
will, when linearized in $\delta B$ and Fourier transformed, yield a dispersion relation for sound-waves of frequency $w$ and wave-number $k$
\begin{equation}
w - \left( \frac{\partial P}{\partial \rho} \right)^{1/2} k + i \left( 
\frac{4\eta}{3 sT}
 \right) k^2 =0
\end{equation}
It is clear that, for the high wavenumber  diffusive mode, the speed of diffusion 
\[\ v = \frac{w}{k} \sim k  \]
 Therefore, for $k \frac{4\eta}{3 sT} \gg 1 $ we go to limit where modes of propagation can be travel faster than light, then principle of causality will be violated.  As long been known \cite{hiscock}, such lack of causality implies lack of stability.  {The Lagrangian treatment in the previous section, with a demonstration the Lagrangian is unbound, confirms the lack of stability is not an artifact of leading-order approximations but a fundamental feature of the theory.}

That Higher-order NS appears neither stable nor causal is not surprising, since the two concepts correlated in relativistic systems, as non-causality generally implies the absence of a ``vacuum'' (either quantum or thermal), which in turn generically leads to instabilities in the effective theory.  The most widely accepted way to solve this issue \cite{israelst} is to {\em promote} $\Pi_{\mu \nu}$  to independent degrees of freedom at second order.  This means to consider the energy-momentum tensor to be
\begin{equation}\label{general}
T^{\mu\nu} = T^{\mu\nu}_0 + \Pi^{\mu\nu}{,} 
\end{equation}
where $T^{\mu \nu}_0$ is given by Eq. \ref{tensor} and $\Pi^{\mu \nu}$ is arbitrary beyond the constraints of symmetry and transversality 
\[  \Pi_{\mu \nu} = \Pi_{\nu \mu} \eqcomma u_\mu \Pi^{\mu \nu}=0{,} \]
the equation of motion is engineered to keep $\Pi^{\mu \nu}$ symmetric and to have the Navier-Stokes as the asymptotic value. {
\begin{equation}\label{IS}
\tau _\pi^\eta \Delta^{\kappa \mu} \Delta^{\zeta \nu} u^\alpha \partial _\alpha  {\pi_{\kappa\zeta}} + {\pi^{\mu\nu}} = \sigma^{\mu \nu}_\eta + \order{(\partial u)^2}{,}
\end{equation}
\begin{equation}\label{ISzeta}
\tau _\pi^\zeta  u^\alpha \partial _\alpha  \Pi + \Pi = \sigma_\zeta + \order{(\partial u)^2}.
\end{equation}
Where, $\sigma^{\mu \nu}_{\eta,\zeta}$ are respectively the Navier-Stokes terms  for shear and bulk viscosity (Eq. \ref{eqns}). {$\Pi^{\mu\nu} = \pi^{\mu\nu} + \Delta^{\mu\nu}\Pi $ where} {$\pi^{\mu \nu}$} is symmetric and transverse and has 5 DoFs \cite{romdev,kovfluct} (the two $\Delta^{...}$ projects make sure  {$\pi^{\mu \nu}$} remains orthogonal to velocity provided initial conditions are so), while { $\Pi \equiv \Pi^\mu_\mu /3 $} { $\Pi = \Pi^\mu_\mu $} is one number representing the trace.}

To understand equation (\ref{IS}) from the EFT Lagrangian approach, we have to remember that $ {\pi_{\mu\nu}}$ get promoted to independent degrees of freedom, subject to the constraint of symmetry and transversality with velocity.   
This degree of freedom is {\em not} a Noether current (and hence cannot be obtained from an Eq. such as \ref{noetherT}) or a conserved quantity.
Hence, the way to write this down in our formalism is to use
\begin{equation}
\label{xdef}
 {\pi_{\mu\nu}} = X_{IJ} \bar{A}^{IJ}_{\mu \nu} \eqcomma \Pi = { X_{IJ}\frac{1}{3} B^{IJ},}
\end{equation}
{where $X_{IJ}$ is a new symmetric matrix of degrees of freedom, one which is not necessarily isotropic but still homogeneous and $\bar{A}^{IJ}_{\mu \nu}$ are traceless and traced parts of the most general rank 2 tensor transversal to flow projected onto the comoving frame, the scalar-tensor decomposition of Eq. \ref{BAdef}
\[\ 
A^{IJ}_{\mu\nu} = \frac{1}{3} \delta_{\mu\nu} \partial_\lambda \phi^I \partial^\lambda \phi^J + \frac{1}{2}( \partial_\mu \phi^I \partial_\nu \phi^J +
\]
\begin{equation}
  \label{aijdef}
 \partial_\nu \phi^I \partial_\mu \phi^J - \frac{2}{3} \delta_{\mu\nu} \partial_\lambda \phi^I \partial^\lambda \phi^J)  .  
\end{equation}}
The shear part therefore is
{
\begin{equation}
\label{shearpi}
 {\pi_{\mu\nu}} = X_{IJ} \frac{1}{2}( \partial_\mu \phi^I \partial_\nu \phi^J + \partial_\nu \phi^I \partial_\mu \phi^J - \frac{2}{3} \delta_{\mu\nu} \partial_\lambda \phi^I \partial^\lambda \phi^J),     
\end{equation}}
and the bulk part is
{
\begin{equation}
  \label{bulkpi}
\Pi = X_{IJ} \frac{1}{3} \partial_\lambda \phi^I \partial^\lambda \phi^J .
\end{equation}
Each of these new degrees of freedom can be doubled, as were equilibrium degrees of freedom in the previous section, to model dissipative dynamics.  The doubled degrees of freedom are denoted by $X_\pm,\Pi_\pm$.
We can see the need for $A^{IJ}_{\mu\nu}$ by applying Eq. \ref{projector}.  It has the same role as Eq. \ref{ns}, but its coefficient is an independent degree of freedom and not fixed by shear viscosity.}

The necessity of new degrees of freedom at second order, somewhat arbitrary in other approaches, becomes clear here from symmetry and causality arguments.   Since $\phi_I$ are already fixed by the initial conditions (up to the volume-preserving and $SO(3)$ diffeomorphism invariance of hydrodynamics), second-order terms in the Lagrangian in therms of just $\phi_I$ would be either non-dissipative, non-causal, or lead to violations of conservation laws: Ostrogradski's theorem \cite{ostro,ostro2} (see appendix \ref{appctp}) prevents us from employing second order derivatives of $\phi$ as degrees of freedom \footnote{Note that the Ostrogradski instability s very different \cite{ostro} from the instability due to the Lagrangian having saddle points which plagues the Navier-Stokes theory (Fig. \ref{action} top panel).  In the Ostrogradski case dynamics is still well-defined, but energy is not positive-definite.   For an isolated system this does not necessarily produce instabilities, but coupled to another system such a system cannot reach equilibrium.  For an unbounded {\em lagrangian} the instability happens in an isolated system too {.} } (as would have been the natural continuation in a gradient expansion).  A $ {\pi_{\mu\nu}}$ dependent on the first derivatives of $\phi_I$, translated into Eq \ref{eqns} can only become of the form
\[\    
 {\pi_{\mu\nu} = f(\partial_\alpha u_\beta). }
\]
As shown in eq \ref{projector}, such a term projected perpendicular to $u_\mu$ will generally contain a pathologically linear $B^{-1}_{IJ}$ term.  At the Lagrangian level, this is the realization that, after linearization, will give a dispersion relation as a complex polynomial
\[\  w(k) = A_n k^n{,} \]
without extreme fine-tuning, such a dispersion relation will inevitably be non-causal.  

We note that, provided the lagrangian contains square terms of $ {\pi_{\mu\nu}}$, eq. \ref{xdef} will contain terms $\sim B_{IJ}^2$ and hence, provided $\Pi_{\mu \nu}$'s normalization is positive-definite, should be stable even if $X_{IJ}$ are arbitrary.  Thus, a bounded Lagrangian stable w.r.t. Ostrogradski's conditions should contain terms at least up to $\sim \Pi_{\mu \nu} \Pi^{\mu \nu}$ where $\Pi_{\mu \nu}$ depends on $X_{IJ}$.  The leading-order dependence compatible with Lorentz symmetries is Eq. \ref{shearpi} and \ref{bulkpi}.

The introduction of these new degrees of freedom in Eq. \ref{xdef} and the absence of additional conserved currents means {  equation \ref{IS} cannot be obtained from a Noether current equation of the form of $\partial_\mu J^\mu = R$ and $J^\mu,R$ are deriveable from the symmetrized coordinates (Eq. \ref{noetherT} and \ref{entropy} are of this form), but must be obtained from the full Lagrangian equation of motion
\begin{equation}
\label{lageq}
\partial_\mu \frac{\partial \mathcal{L}}{\partial (\partial_\mu Z)} =  \frac{\partial \mathcal{L}}{\partial Z} 
\end{equation}
where  $Z = \Pi_{\mu \nu}(X_{IJ},\phi_I)$.  The right hand side of this equation is fixed by the necessity of the asymptotic theory to relax to Navier-Stokes
by symmetry, since $\Pi_{\mu \nu}$ is perpendicular to $u_{\mu}$, the new DoFs have to have the same number of elements as $A^{IJ}_{\mu\nu}$, $B_{IJ}$, hence the necessity for the new matrix.}

Using $\Pi_{\mu \nu}$ as variables for the Lagrangian allows us to more readily make contact with standard I-S equations, but obscures the role of symmetries of the $X$ and $A$ terms.
In particular, note that the projected traceless part of $\frac{\partial \mathcal{L}}{\partial (\partial_\alpha A^{IJ}_{\mu \nu})} \partial_\beta A^{IJ}_{\mu \nu} \sim \Pi_{\alpha \beta} \partial_\gamma u^\gamma$, the residual violation of conformal symmetry within volume preserving diffeomorphism invariance for $\Pi_{\mu \nu}$ (the third term of Eq. 3.12 of  \cite{romconf}).  A conformal transformation on the Lagrangian would generate such a term in the energy momentum tensor, and hence this term must appear with the opposite sign if conformal invariance was enforced at the IS level.  A consistent development of conformally invariant hydrodynamics is however left for a different work, since, as can be seen from these equations, the relationship between $X_{IJ},\Pi_{\mu \nu}$ and $A$ are complex enough to require some work for conformal symmetry to be fully implemented (there is much more to a conformal fluid than the absence of bulk viscosity){.}

The second law of thermodynamics prevents $X_{IJ}$ from having non-dissipative terms and requires, for long distances, that $\Pi_{\mu \nu}$ relaxes to its Navier-Stokes value.   The simplest Lagrangian of this form is (note that the projection of $\Pi_{\mu \nu}$ perpendicularly to $u_\mu$ is taken when {$\Pi_{\mu \nu}$} is defined) 

{
\begin{equation}
\label{isl}
 \mathcal{L} = \mathcal{L}_{ideal}(B_{\pm}) + \mathcal{L}_{IS-shear} + \mathcal{L}_{IS-bulk} + \mathcal{L}_2 \left( \left(\partial \phi_{\pm}\right)^2  \right){,}
\end{equation}}
\[\ \mathcal{L}_{IS-shear}= \frac{1}{2}\tau_\pi^\eta \left({\pi^{\mu\nu}_-} u^\alpha_+ \partial_\alpha {\pi_{\mu\nu +}} - {\pi^{\mu\nu}_+} u^\alpha_- \partial_\alpha {\pi_{\mu\nu -}} \right)+ \]
\[\ + \frac{1}{2} {\pi^{\mu\nu}_{\pm}} {\pi_{\mu\nu \pm}} +  \underbrace{\left[ \left( A^{\circ} \right)_{\mu \nu}^{IJ} \partial^\mu K^\nu \right]_{\pm}}_{\sim \sigma_{\mu \nu}^\eta}{,}  \]

\[\ \mathcal{L}_{IS-bulk}= \frac{1}{2}\tau_\pi^\zeta \left(\Pi_- u^\alpha_+ \partial_\alpha \Pi_{ +} -  \Pi_+ u^\alpha_- \partial_\alpha \Pi_{ -}\right)+ \]
\[\   + \frac{1}{2} \Pi_{\pm}^2  + \underbrace{\left[  K_\mu \partial^\mu B \right]_{\pm}}_{\sim \sigma^\zeta}{.}  \]
{Where $\left( A^{\circ} \right)_{\mu \nu}^{IJ} \equiv A^{IJ}_{\mu\nu} / B_{IJ} $ and }the first term is the ideal one (Eq. \ref{idealhy}), the next two terms give the dissipative dynamics of $\Pi_{\mu \nu}$ and the last one gives the Navier-Stokes source Eq. \ref{eqns} (note the explicit Lagrangian dependence of section \ref{secns} disappears, as its presence would give rise to unstable modes).
$\mathcal{L}_2$ contains non-dissipative ``hydrodynamic'' terms to second order in Gradient enumerated in works such as \cite{secondorder}.  It will contain shear and bulk-mixing terms, as well as further restrictions due to, for example, conformal symmetry. Its construction, as in all EFTs, is based on enumerating all second order terms conpatible with Lorentz  and homogeneity symmetries{.}

It is easy to see that Eq. \ref{IS} arises as an equation of motion w.r.t. $X$ and a 

\[\ T^{\mu \nu} = T_0^{\mu \nu}(B,u^\mu) + \Pi^{\mu \nu}(X_{IJ},A^{\mu \nu}){,}\]
where $T_0^{\mu \nu}$ is given by the ideal tensor Eq. \ref{noetherT} and $\Pi^{\mu \nu}$ is perpendicular to $u_\mu$.
It is also clear that this Lagrangian is stable (the action is bounded, with one minimum, in the near-hydrostatic ($\phi_I \simeq \vec{x}$) limit, as shown in Fig. \ref{action} bottom panel) against both the ``old'' DoFs $\phi_I$ and the new one $X_{IJ}$: $\Pi_{\mu \nu} \Pi^{\mu \nu}$ is positive definite, and, as long as $\tau_\pi$ is large enough to guarantee slower than light diffusion propagation the kinetic term will not contain ghost modes.  {Beyond this restriction, $\tau_\pi^{\eta,\zeta}$ can have an arbitrary dependence on the equilibrium DoF $B${.}

\section{Discussion \label{secfinal}}

Our results provide a general method to extend the EFT which completes the insights of \cite{marrochio,shipu,denicol}.  For stability the theory in its Lagrangian description has to contain strictly positive powers of $B_{IJ}$ and its derivatives, which means even powers with positive combinations of gradients and traces at the lagrangian level.   An odd power will generally mean an unstable equation of motion, of the type found in \cite{shipu,marrochio}.

Beyond these constraints, all the allowed combinations of $B,B_{IJ},X_{IJ}$ and their gradients can go into the Lagrangian, with terms suppressed in powers of the gradient by the Knudsen number.   Naively, terms with $X_{IJ}$ are ``one higher order in gradient'' w.r.t. $B_{IJ},B$.  This is what one expects when the theory is close to equilibrium and the viscosity $\eta$, relaxation time $\tau_\pi$, temperature $T$ and and entropy $s$ are related by $\tau_\pi T \sim \eta/s$.  This relation however is {\em not} universal.  In situations of critical slowing down it manifestly fails \cite{critslow}, since shear viscosity is at a minimum while $X_{IJ}$ never equilibrate.

More generally, since $X_{IJ}$ are independent degrees of freedom, their equilibration time in a classical theory depends on initial conditions, a dependence that, at least in examples with reduced dimension, has been shown to be highly non-trivial \cite{mauricio,mepratt}.  This is why the proposal of \cite{reynolds} to treat such terms as independent of the gradient expansion appears justified.
That said, the fact that the leading order dynamics of such terms is only dissipative does provide an explanation for the applicability of the EFT expansion in a regime where there is scope for it failing \cite{eftfail}.  After all, hydrodynamic behavior appears universal in nature.

These extra degrees of freedom provide a conceptual framework for unifying Israel-Stewart hydrodynamics with anisotropic hydrodynamics, recently developed as a successful phenomenological theory \cite{aniso1,aniso2}.  There is in fact no distinction between usual Israel-Stewart and anisotropic hydrodynamics in principle, since the appearance of DoFs corresponding to anisotropies is inevitable in the former.  The phenomenological success of \cite{aniso1,aniso2} might be due to the fact that the latter keeps track, within the gradient expansion, the fact that anisotropies in the gradient are very different in the transverse and longitudinal direction.  Hence, the number of terms considered should vary between these directions.

It should be remembered that the dissipative terms generally violate conservation laws \cite{galley}, but, for an EFT expansion of a highly symmetric theory, this violation typically goes as a higher order than the Lagrangian \cite{grozdanov}.  The lagrangians examined here are no exception, since the violation of energy conservation examined in Section \ref{secns} is $\sim (l_{mfp}\nabla)^2$ \cite{grozdanov}{ while that in \ref{secis} should be, from \cite{galley} Eq. 2.63
\[\ \Delta T^{0i} \sim  \dot{X} \frac{\partial \mathcal{L}_{diss}}{\partial X} + \ddot{X} \frac{\partial \mathcal{L}_{diss}}{\partial \dot{X}} + (l_{mfp}\nabla)^2(\nabla X)  \]
\[\ \sim (l_{mfp}\nabla)(\nabla X)^2 +  (l_{mfp}\nabla)^2(\nabla X){.} \]}
However, as the convergence of the IS expansion is unclear in the general case, this issue will need to be examined on a case-by case basis.

The introduction of $X_{IJ}$ at second order raises the question of weather additional degrees of freedom of this sort are required at higher order. To answer this question, one must remember that Ostrogradski's theorem can be applied sequentially.   Once second order derivatives of $X_{IJ}$ are taken into account, new degrees of freedom are required to ensure thermodynamic stability of the theory. These terms cannot be rank two tensors, since  by the Coleman-Mandula \cite{cm} theorem the only conserved quantity is $T_{\mu \nu}$ and hence a hypothetical $X^{IJ}_2$ can be always absorbed into a redefinition of the non-conserved $X_{IJ}$.   Hence, new terms will come as contractions of higher rank tensors, for example $X^{IJKL}_2 X_{IJ}X_{KL}$, with new higher rank terms being again non-conserved. Within the Boltzmann equation \cite{reynolds} one can consider these extra terms as related to moments of the Boltzmann equation, but our theory is ''bottom-up'' so, beyond these terms encoding microscopic correlations of some sort \cite{meeft}, we can say nothing about them except their symmetry properties.

A topic which we did not elaborate is the role of the entropy in this dynamics. The conservation of the entropy current arises as an equation of motion if the non-dissipative Lagrangian in section \ref{secideal} is analyzed in terms of its equations of motion rather than the energy-momentum tensor. To investigate the entropy current systematically in terms of the Lagrangian, we would need a quantitative relation between action and entropy. In the adiabatic limit (microscopic DoFs are parametrically faster than macroscopic ones) we can use Matsubara's prescription \cite{kapustagale} 
\begin{equation}
t \rightarrow \frac{i}{T} \eqcomma S_{quantum} \rightarrow \frac{F}{T}{,}
\end{equation}
(where $S_{quantum}$ is the action of the quantum theory, $F$ is the free energy and $T$ the temperature)
together with the semiclassical limit ($T_0 \rightarrow \infty$. Note that expansion beyond this limit is uncorrelated from the gradient expansion \cite{mevisc,burch}. It generally goes as $T_0 \sim g^{-1}$, so $\sim N_c^{-2}$ in $SU(N_c)$ Gauge theories)
\begin{equation}
\label{semiclassical}
F=T \ln Z \eqcomma Z \simeq \exp\left(T_0^4 S_{minimum} \right){,}
\end{equation}
and the definition of the CTP formalism \cite{grozdanov} of Eq. \ref{ctp} it is clear that the macroscopic entropy of a fluid at temperature $T$ will be
\begin{equation}
 s = \frac{1}{T}  \int_0^T dT' \int_0^{2\pi}  \mathcal{L}\left(\phi_{\pm}(x,t=e^{i\theta/T'}  \right) d\theta d^d x {.}
\end{equation}
These relations ensure that there is a one to one correspondence between the lagrangian terms in the effective theory and the entropy current, which needs to be investigated term by term.

In the ideal limit, where the mixing term in Eq. \ref{ctpdef} vanishes, this entropy is simply equal to {$T_0^3 B^{1/2}$}  and is conserved, but, as elucidated in \cite{galley}, dissipative terms generally introduce ``violations'' of Noether's theorem via sources in the unobserved part of the system.  Entropy conservation is of course the most well-known of such violations.  For $T \ll T_0$ we should recover the entropy formulae in \cite{grozdanov}. 
Hence, the formulae argued for in \cite{romdev}
{\begin{equation}
\partial_\mu (s u^\mu) = \partial_\mu (\sqrt{B} u^\mu) \sim \left. 0\right|_{i=0} + \left. \Pi_{\mu \nu} \Pi^{\mu \nu} \right|_{i \ge 1}{,}
\end{equation}
where i corresponds to gradient order }
can be justified from the Lagrangian description.  Higher order terms in the effective Lagrangian with couplings between $X_{IJ}$ and gradients of $B$ will therefore also enter in the entropy current.   In a sense, the non-decrease of entropy in this formulation is equivalent to the Lagrangian being bound from below, which reinforces the connection between our work and \cite{shipu,marrochio}{.}

This ``semiclassical'' limit, though, is likely to severely underestimate entropy formation at {\em low} viscosity, for the first of the terms in the inequality required for hydrodynamics to hold \cite{betz,burch,mevisc}
\begin{equation}
\label{hyerarchy}
 \frac{1}{T_0} \ll \frac{\eta}{Ts} \ll \frac{1}{\partial u}{,}  
\end{equation}
will inevitably break down as $\eta \rightarrow 0$.  In this case Eq. \ref{semiclassical} will also break down and one will have to do the full functional integral, with $T_0$ as the coupling constant, ie replace in Eq. \ref{semiclassical}
\[\  e^{S_{minimum}(B,X)} \rightarrow \int \mathcal{D} \left[B,X\right] e^{T_0^4 \int \mathcal{L}(B,X) d^4 x}{,}  \]
Relatively elementary considerations, such as the fact that vortices, which require no energy to form, do not propagate suggests the expansion around $T_0$ is hihgly non-perturbative, possibly necessitating nomerical methods to be properly taken into account. {Physically, this means that excitations of macroscopic degrees of freedon $\phi_I$ and $X$, and coupling of these to microscopic degrees of freedom, gives a non-negligible description of the entropy{.} }
{The most likely appearanche of such degrees of freedom is within the variables $A_{IJ}^{\mu \nu}$ of Eq. \ref{aijdef}, classically forbidden within the ideal hydrodynamic limit, as shown by the first efforts to simulate this system numerically \cite{burch}. }

Such effects might be crucial in the low viscosity high flow limit, where the phenomenon of turbulence \cite{landau} occurs in the classical limit.
In our picture, turbulence can be understood as the occurrence of multiple irregularly distributed minima and saddle points in the action as a function of field configurations (Fig. \ref{turbo}).   In this limit deviations from the semiclassical approximations are likely to dominate even for ``small gradients'', precisely because $T_0 \sim \ave{\partial u}$, and the EFT expansion likely breaks down \cite{eftfail,meeft}. 

Such dynamics might have a role in clarifying the mysteries the phenomenon of turbulence still yields.   Since $T_0$ in equation \ref{hyerarchy} is proportional to the distance scale of {\em microscopic} DoFs, $\order{1000(100fm^3)^{-1}}$ in a heavy ion collision (or $10^4/m^3$ in a cold atom system), such corrections, as yet completely unexplored, could become crucial to connect our EFT to phenomenology.  Naively, the fluctuation-driven mixing of turbulent and microscopic degrees of freedom will shorten microscopic thermalization, invalidating ``hydrodinamicization'' scenarios such as \cite{chesler}, at least beyond the planar limit.
\begin{figure}[t]
  \includegraphics[height=0.2\textheight]{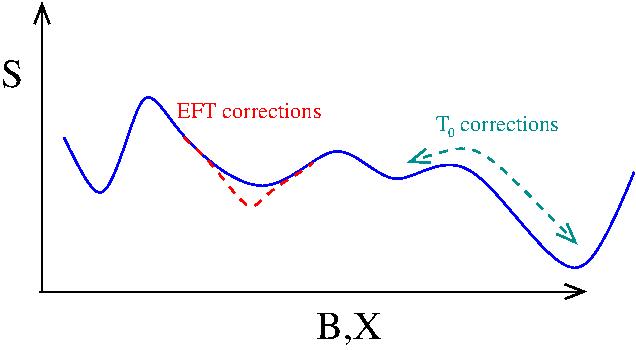}
  \caption{\label{turbo}The profile of the action in a "turbulent" system (small viscosity and well away from the hydrostatic limit), and a qualitative sketch of the role of corrections of gradient and $T_0$ terms{.}
}
\end{figure}
Effective field theories emerge from renormalization group flow of microscopic theories.   Hydrodynamics is ultimately {\em not} different, except for the fact that the effective Lagrangian must be non-unitary to take into account entropy exchanges between the microscopic and macroscopic scales \cite{meeft}.  Furthermore, this exchange assumes the existence of a microscopic scale which is not invariant under macroscopic transformations, requiring terms non-invariant under rescaling even in a theory which macroscopically is invariant under such transformations.This is manifest at first order already, since viscosity, ultimately due to a "physical" microscopic scale, is represented by terms that violate rescaling diffeomorphisms. {Given these considerations, any strongly-coupled relativistic microscopic field theory should in principle coarse-grain to something like Eq. \ref{Hyexpansion2} provided coarse-graining is done within the CTP formalism, as suggested in \cite{polctp}{.} }

In conclusion, we have shown that the Lagrangian formalism incorporates naturally both the Navier-Stokes and the Israel-Stewart terms of hydrodynamics, and naturally explains the appearance and scaling of these terms in a way that is somewhat model-dependent and ad hoc in other approaches.   We await to see to what extent this proof can be generalized to arbitrary order.

\textit{Acknowledgements} GT acknowledges support from FAPESP proc. 2014/13120-7 and CNPQ bolsa de produtividade 301996/2014-8.    DM would like to acknowledge CNPQ graduate fellowship n. 147435/2014-5
We would like to thank Jorge Noronha, Saso Grozdanov and Jun Takahashi for discussions and suggestions.
\appendix
\section{A Lagrangian description of conservative systems \label{appctp}}
\subsection{Introduction}
In this short review we use the notation of \cite{grozdanov}.
The Lagrangian formulation is a very powerful technique for formulating both classical and quantum problems. There is a demand to systematic procedure of incorporate general dissipative features in variational principle. The best way to establish this change is by theoretical modifying the formulation of the principle of least action \cite{caltech,morse,galley1}. The main point that variational principle needs \textit{boundary conditions} and temporal invariance. As a result Sturm-Liouville's theorem implies time-symmetry, something easy to derive by considering Greens functions \cite{galley}.

To go beyond this limitation 
we will use Closed-Time-Path (CTP) \cite{grozdanov} in the description of a closed system.  For each field $\psi(x)$, we double to a set of variable $\psi \rightarrow ({\psi^+},{\psi^-})$ with same initial conditions. Physically, in the context of Lagrangian mechanics shows each variable represent an equation of motion where one "absorb" and another "loses" energy.  
\begin{equation}
\label{ctpdef}
S_{CTP}[\hat{\psi}]=  \int_{t_f}^{t_i}d^{d+1}x\left \{ \mathcal{L}_s[\psi ^+] - \mathcal{L}_s^*[\psi ^-]\right \}.
\end{equation}
In the context of this work, the two systems can be considered to be the infrared (macroscopic) and ultraviolet (microscopic) degrees of freedom.

To reproduce dissipative dynamics via a Lagrangian, each DoFs has to ensure the \textit{equality conditions} (sometimes called "physical limit"). In other words, $\psi^+ (t_f, $\textbf{x}$) = \psi^- (t_f,$ \textbf{x}$)$, and $\partial^n_t \psi^+(t_i,$\textbf{x}$)=\partial^n _t \psi^-(t_i,$\textbf{x}$)$, with $\psi^\pm$ an function of class n. \textit{initial conditions} maintains a causality in non-conservative system while generating a difference between advanced and retarded Greens functions.

The CTP has a degeneracy associated  with $\psi^+ \leftrightarrow \psi^- $, that is, both axes obey the same equation of motion, due to the double DoFs what imply in the action \ref{ctpdef},
{
\begin{equation}
  S_{CTP}[\psi^+,\psi^-]=-S^*_{CTP}[\psi^-,\psi^+]{.}
\end{equation}}

The degeneracy introduced in $\psi$ disappear in physics limit (p.l.), $\psi^+(x)=\psi^-(x)$, only the variable $\psi^+$ has physical meaning and another vanishes \cite{grozdanov}.  The effective Lagrangian is therefore of the form (in our case $\psi=\phi$, introduced in section \ref{secideal} and $X$ in section \ref{secis})
\begin{equation}\label{ctp}
S_{eff}[\hat{\psi}] = S_s[\hat{\psi^+}] - S^*_s[\hat{\psi^-}] + S_i[\hat{\psi^+,\psi^-}]{,}
\end{equation}
where the non-accessible DoFs $S_i\left(\hat{\psi}\right)$  are microscopic or {``high energy''}, called internal energy and assumed to be a perturbation close to the ideal limit. The subtraction is equivalent \cite{galley1,grozdanov} to integrating out unseen degrees of freedom.

The variational principle of $S_s$ is defined by the Hessian matrix $\delta ^2 S_i / \delta \phi^+ \delta \phi^-$. The first and second term are responsible to conserved current of energy-momentum. The $S_i$ is a small perturbation, where contains all terms due to integration out more dissipative forces.  To ensure Lioville's theorem (unitarity in the quantum theory), we would need to compute all terms of action including the ones we do not keep track of.
Care needs to be taken to translate standard Lagrangian mechanics results into the doubled coordinate formalism.  For instance, as discussed in \cite{galley} dissipative terms $S_i(\psi_+,\psi_-)$ generally ``break'' conservation laws inferred from Noether's theorem (friction is an everyday example).  Nevertheless, it is possible to extend such theorems into the dissipative domain and gauge their applicability on a case by case basis.    In the next section we will consider the case of Ostrogradski's theorem, necessary for the derivation in section \ref{secis}.   

Note that the approach described here is not unique.   For instance, one can integrate out the unseen degrees of freedom explicitly, as was done in \cite{porto}.  We think the method used here has more potential for a systematic gradient expansion, since, as shown in \cite{grozdanov}, violations of non-entropy conservation laws (energy, charge, etc) can be systematically organized as ``higher order gradient'' terms, by ensuring the correct gradient power-counting for $S_i$ w.r.t. $S_s$.  
In a direct integrating out procedure, all conservation law violations are parametrically similar (essentially the ``violation'' is contained in the DoFs one integrates out), and because of this the approximate invariance under conservation laws in \cite{porto} depends on linearization as well as the EFT expansion.
\subsection{Ostrogradski's theorem}

Ostrogradski's theorem \cite{ostro,ostro2} limits ``well-behaved'' theories to Lagrangians with two derivatives.   Higher derivative terms, even with a well-defined lagrangian minimum, will have an unstable mode in the {\em Hamiltonian} (essentially, negative energy ``states'').   This makes local thermalization for such systems {\em and any systems coupled to them} obviously problematic.  In particular, the EFT expansion generally breaks down because the system has a vacuum instability.

To check that Ostrogradski's theorem applies to dissipative Lagrangians consider a doubled field Lagrangian density with field coordinate $\phi(x^\mu)$
\[
\Lambda(\phi^{\pm}, \phi^{\pm}_{\mu} , ... , \phi^{\pm}_{\mu_1 ... \mu_n}) =  \mathcal{L}_o(\phi^{+}, \phi^{+}_{\mu} , ... , \phi^{+}_{\mu_1 ... \mu_n})
\]
\begin{equation}
+ \mathcal{K}(\phi^{\pm}, \phi^{\pm}_{\mu} , ... , \phi^{\pm}_{\mu_1 ...  \mu_n}).
\end{equation}
Where $\phi^{\pm}_{\mu_1 ... \mu_n}  \equiv \partial_{\mu_1} ... \partial_{\mu_n} \phi^{\pm} $. We will use $(\phi^{\pm}, \phi^{\pm}_{\mu} , ... , \phi^{\pm}_{\mu_1 ... \mu_n}) \equiv (\phi^{\pm}_{\alpha_n})$, $ \alpha $ is a set formed by $\left\{ \mu, ... , \mu_1 ... \mu_n \right\}$, with $ 0 \leq n \leq m  $ and  $m$ is degree of Lagrangian. Assuming \textit{non degeneracy} we can invert the highest derivative function as $\phi^{\pm}_{\alpha_m} = ( \phi^{\pm}_{\alpha_l}; \pi^{\pm  \alpha_m} )$ with $0 \leq l \leq m-1$. 

The canonical momentum is defined as $
\pi^{\pm \alpha_n} \equiv \frac{L}{\partial \phi^{ \mp \alpha_n}}$ when apply physical limit ("p.l.") just $\pi^{+ \alpha_n} \equiv \frac{L}{\partial \phi^{ - \alpha_n}}$ generalizing momentum is straightforwardly. The Hamiltonian is
\[
\mathcal{H}(\phi^{\pm}_{\alpha_n}) =  \phi^{i_{1}}_\mu \pi^{j_{1} \alpha_1} + ... +  \phi^{i_{n-1}}_{\alpha_{n-1}} \pi^{j_1 \alpha_{n-1}} 
\]
\begin{equation}\label{hami2}
+  \bar{\phi}^{i_n}_{\alpha_n} \pi^{j_1 \alpha_n} - \Lambda (\phi^{\pm}_{\alpha_n}).
\end{equation}
The set $(i_n,j_n)$ =  $\left\{ ( +,- ) ; ( -,+) \right\} $,  $0 \leq n \leq m $  . CTP is carried over a new poison bracket where $f$ and $g$ are functions of coordinate and momentum. 
\[
\left\{ \left\{ f,g \right\} \right\} = 
\bigg\{ \frac{\partial f}{\partial \phi^{+}_{\alpha_{n-1}}}\frac{\partial g}{\partial \pi^{-}_{\alpha_n}} - \frac{\partial f}{\partial \phi^{+}_{\alpha_{n-1}}}\frac{\partial g}{\partial \pi^{-}_{\alpha_n}} \bigg\} 
\]
\begin{equation} 
+ \bigg\{ \frac{\partial f}{\partial \phi^{-}_{\alpha_{n-1}}}\frac{\partial g}{\partial \pi^{+}_{\alpha_n}} - \frac{\partial f}{\partial \phi^{-}_{\alpha_{n-1}}}\frac{\partial g}{\partial \pi^{+}_{\alpha_n}}  \bigg\}{.}
\end{equation}
It is straightforward to show the new equation of motion to this Hamiltonian is
\[
\partial_{\nu} \phi_{\alpha_{n-1}} = \frac{\partial \mathcal{H}}{\partial \pi^{\nu \alpha_{n-1} }} - \left [\frac{\partial \mathcal{K}}{\partial \pi^{\nu \alpha_{n-1}}_- } \right]_{p.l.} = 
\]
\begin{equation}
\left\{ \phi_{\alpha_{n-1}} ,\mathcal{H} \right\} -  \left\{ \left\{  \phi_{\alpha_{n-1}}^{-} ,\mathcal{K} _{p.l.} \right\} \right\}_{p.l.}{,}
\end{equation}
\[
\partial_\nu \pi^{\nu \alpha_{n-1} } = - \frac{\partial \mathcal{H}}{\partial \phi_{\alpha_{n-1}} } + \left [ \frac{\partial \mathcal{K}}{\partial \phi^-_{\alpha_{n-1}} } \right]_{p.l.} = 
\]
\begin{equation}
\left\{ \pi^{\alpha_{n}} , \mathcal{H} \right\} - \left\{ \left\{ \pi^{\alpha_{n}}_- ,\mathcal{K} \right\} \right\}_{p.l.}{.}
\end{equation}
One can see, the first term lhs is \ref{hami2} and second represent the coupling between the two doubled degrees of freedom $(\phi^+,\phi^-)$ which corresponds to a holonomic force, between system and environment. Provided the new Hessian matrix is invertible
\begin{equation}
 \frac{\delta \mathcal{L}}{\delta \phi^+ \delta \phi^-} \neq 0{,}
\end{equation}
 the instability described in \cite{ostro,ostro2}, and all the problems inherent to coupling this system to other systems, will also appear in the dissipative Lagrangian.
 
 \begin{equation}\label{3}
 \bigg [ \frac{d}{dt}  \frac{\partial \Lambda}{\partial \dot{q}_- }  - \frac{\partial \Lambda}{\partial q_-} \bigg ]_{f.l.} = 0{.}
\end{equation}

\end{document}